\documentclass[useAMS,usenatbib]{mn2e}

\usepackage{graphicx}

\title[Amplitude Saturation in $\beta$~Cephei Models]{Amplitude Saturation in $\beta$~Cephei Models}
\author[R. Smolec and P. Moskalik]{R. Smolec$^{1}$\thanks{E-mail:
smolec@camk.edu.pl (RS); pam@camk.edu.pl (PM)} and P. Moskalik$^{1}$\footnotemark[1]\\
$^{1}$Copernicus Astronomical Centre, Bartycka 18, 00-716 Warsaw, Poland}
\begin{document}

\date{Accepted . Received ; in original form }

\pagerange{\pageref{firstpage}--\pageref{lastpage}} \pubyear{2002}

\maketitle

\label{firstpage}

\begin{abstract}

Although the driving mechanism acting in $\beta$~Cephei pulsators
is well known (e.g.~\citealt{WDAP93}), problems concerning
identification of amplitude limitation mechanism and non-uniform
filling of the theoretical instability strip, remain to be solved.
In the present analysis, these problems are addressed by
non-linear modelling of radial pulsations of these stars. In this
approach radial modes are treated as representative for all
acoustic oscillations.

Several models of different masses and metallicities were
converged to limit cycles through \citet{RS74} relaxation
technique. Resulting peak-to-peak amplitudes are of order of
$\Delta V=0.3$\thinspace mag. Such amplitudes are significantly
larger than those observed in $\beta$~Cephei pulsators. Assuming
that all acoustic modes are similar, we show that {\em collective
saturation} of the driving mechanism by several acoustic modes can
easily lower predicted saturation amplitudes to the observed
level. Our calculations predict significant decrease of saturation
amplitudes as we go to high mass/high luminosity models. However,
this effect is not strong enough to explain scarcity of high mass
$\beta$~Cephei variables. A possible weakness of the collective
saturation scenario is that estimated line-broadening, resulting
from excitation of many high-$l$ modes, might be higher than
observed in some of the $\beta$~Cephei stars. We argue that this
difficulty can be overcome by allowing g-modes to participate in
the saturation process.

We also discuss robust double-mode behaviour, encountered in our
radiative models. On a single evolutionary track we identify two
double-mode domains with two different mechanisms resposible for
double-mode behaviour. The non-resonant double-mode domain
separates first overtone and fundamental mode pulsation domains.
The resonant double-mode domain appears in the middle of the first
overtone pulsation domain. Its origin can be traced to the
$2\omega_1=\omega_0+\omega_2$ parametric resonance, which
destabilizes the first overtone limit cycle.

\end{abstract}


\section{Introduction}

$\beta$~Cephei stars are main sequence early B-type pulsators. Their radial and non-radial p-modes and low-order g-modes are driven through the classical $\kappa$-mechanism acting in the metal opacity bump \citep{PMWD92,WDAP93,AGHS93}. Most of $\beta$~Cephei stars are multiperiodic with typical periods ranging from 3 to 7 hours, and peak-to-peak {\em V}-band amplitudes usually below 60\thinspace mmag. For observational review of astrophysical properties of $\beta$~Cephei stars see \citet{CSMJ93} and \citet{ASGH05}. The latter paper contains the catalogue of 93 confirmed galactic $\beta$~Cephei pulsators. This list should be supplemented by 19 new objects found in ASAS-3 survey data \citep{AP05,GH05}. Most of these newly discovered objects have high amplitudes in comparison with previously known variables (cf.~Table~\ref{tabamp}). This is a selection effect due to automatic algorithm used in the ASAS survey to extract variable objects from several million stars being monitored. Only stars showing an excessive dispersion are subject do detailed periodogram analysis \citep{GP02}. Thus, low amplitude pulsators are not present in the ASAS catalogues. The ongoing periodogram analysis of all the ASAS-3 stars of proper spectral types at least triples the number of detected $\beta$~Cephei pulsators \citep{AP07}. Several $\beta$~Cephei variables were also found in the Magellanic Clouds \citep{APZK02,ZKea04b,ZKea06}.

Although the driving mechanism acting in $\beta$~Cephei stars is well known, several problems remain to be solved. These concern observational boundaries of $\beta$~Cephei instability strip and question of amplitude limitation mechanism. The theoretically predicted instability strip is not completely filled with stars (e.g. \citealt{AP99}, Fig.~\ref{trak}, this paper). Particularly, linear analysis predicts unstable p-modes in slightly evolved massive O-type stars. Despite of observational effort \citep{APZK98}, until very recently we have not known any firm pulsator of this kind. Three candidates were found only recently \citep{AP07}. Also, lack of stars at low-mass end of the theoretical instability strip remains to be explained. Considering detected frequencies, about 38 per cent of galactic $\beta$~Cephei pulsators are monoperiodic. For the purpose of further comparison, in Table~1 we have listed the monoperiodic variables for wich {\it V}-band amplitudes are available. Selection of monoperiodic variables was based on \citet{ASGH05} catalog and publications on ASAS variables \citep{AP05,GH05}. Peak-to-peak amplitudes of these variables are taken from various publications (references are given below the Table). The Table contains both field and cluster variables and is based on analysis of observational data of differing length and quality. Obviously, monoperiodicity of variables listed in Table~1 is a statement reflecting our current knowledge.

Although significant fraction of $\beta$~Cephei variables appears to be monoperiodic, it is very likely, that almost all of the $\beta$~Cephei stars are in fact multiperiodic, with amplitudes of additional modes below current detection limit \citep{ASGH05}. This hypothesis is supported by intensive observation campaigns of individual variables, leading to detection of more pulsation modes with decreasing detection threshold (e.g. \citealt{MJea05}). We also note, that two stars claimed as monoperiodic in \citet{ASGH05} catalogue, are now confirmed multimode pulsators ($\delta$ Cet - \citealt{CAea06}; HD~164340 - \citealt{AP05}). From observational point of view \citet{JDEN05} showed that multiperiodic variables have a larger metal abundance in the photosphere than monoperiodic ones. Linear calculations also show that for higher metal abundance we get more unstable modes \citep{WDAP93}.

Non-linear effects are responsible for the amplitude limitation and thus, non-linear calculations may shed light on problems concerning observed instability strip and instability saturation. So far full non-linear calculations can be done for radial modes only. Radial modes are not rare in the observed $\beta$~Cephei stars \citep{CAPDC03}. Except of BW~Vulpeculae (which is an extreme case), radial modes have similar amplitudes to non-radial modes. Thus, we assume that a radial mode is a typical representative of all acoustic modes. This approximation is applicable to p-modes only and not to g-modes as their physicall nature and thus saturation properties are different.

First non-linear models of $\beta$~Cephei variables were calculated for BW~Vul \citep{WPAC80,PMRB94,AFea04}. This monoperiodic pulsator exhibits most extreme light variability among known $\beta$~Cephei stars, with peak-to-peak amplitude approximately 0.2\thinspace mag in {\em V}. \citeauthor{PMRB94} concluded that BW~Vul is radial, fundamental mode pulsator, which was further confirmed by \citet{CAea95}. \citeauthor{PMRB94} were able to reproduce the observed amplitude, which leads to conclusion that in this case instability may be saturated by a single mode. However, BW~Vul is an extreme case, since most $\beta$~Cephei stars have much smaller amplitudes. For example $\gamma$ Peg is a monoperiodic radial mode pulsator \citep{CAPDC03} with {\em V}-band amplitude of 17\thinspace mmag (Table~\ref{tabamp}). In the present analysis, conclusions about instability saturation are drawn through comparison of calculated non-linear limit cycle amplitudes of $\beta$~Cephei models with observed amplitudes of $\beta$~Cephei stars. Our preliminary results were presented at the {\it Vienna Workshop on the Future of Asteroseismology} \citep{RSPM07}.

This paper is organised as follows. In Section~2 we present our hydrocodes and model construction and analysis. In Section~3 we discuss sensitivity of calculated amplitudes to several numerical parameters (artificial viscosity, numerical mesh). Section~4 contains main results of this paper concerning mode saturation. In Section~5 interesting case of double-mode pulsation in non-convective stellar models is discussed. Conclusions are summarized in Section~6.

\begin{table}
\caption{Monoperiodic $\beta$~Cephei variables, for which {\em V}-light amplitudes are available. Last seven stars are new variables detected in ASAS-3 catalogue. Amplitudes are peak-to-peak values.}
\begin{tabular}{cccc}
Star                 & period [d]& $A_V$ [mag] &  Ref. \\
\hline
$\gamma$ Peg         &   0.1518  &  0.017  &    1 \\
NGC 663 v4           &   0.1940  &  0.040  &    2 \\
NGC 869 v692         &   0.1717  &  0.019  &    3 \\
NGC 869 v992         &   0.1326  &  0.003  &    3 \\
NGC 884 v2299        &   0.3190  &  0.016  &    4 \\
4 CMa                &   0.2096  &  0.034  &    1 \\
KK Vel               &   0.2157  &  0.050   &   5 \\
NGC 4755 v301        &   0.1305  &  0.010  &    6 \\
NGC 4755 v201        &   0.1825  &  0.010  &    6 \\
NGC 4755 v210        &   0.0933  &  0.007  &    6 \\
$\tau^1$ Lup                &   0.1774  &  0.027   &   1 \\
$\delta$ Lup         &   0.1655  &  0.0035  &   7 \\
v1449 Aql            &   0.1822  &  0.074   &   8 \\
NGC 6910 v14         &   0.1904  &  0.017   &   9 \\
NGC 6910 v24         &   0.1430  &  0.009   &   9 \\
BW Vul               &   0.2010  &  0.2  &      1 \\
HN Aqr               &   0.1523  &  0.032   &   10 \\
\hline
CPD$-$62$^\circ$2707 &   0.1417  &  0.110   &   8  \\
HD 133823            &   0.1760  &  0.106   &   8  \\
CPD$-$46$^\circ$8213 &   0.2242  &  0.129   &   8  \\
HD 328906            &   0.1776  &  0.084   &   8  \\
HD 152477            &   0.2650  &  0.071  &    8  \\
BD$-$14$^\circ$5047  &   0.2402  &  0.088   &   8  \\
HD 100495            &   0.1685  &  0.037  &    11 \\
\hline
\end{tabular}\\

 Periods are taken from \citet{ASGH05}. Amplitude references: (1) \citealt{JLMA78}; (2) \citealt{APea01}; (3) \citealt{JKea99}; (4) \citealt{JKAP97}; (5) \citealt{AC82}; (6) \citealt{ASea02}; (7) \citealt{RS72}; (8) \citealt{AP05}; (9) \citealt{ZKea04a}; (10) \citealt{DKFvW90}; (11) \citealt{GH05}.\\{\it Notes on individual stars:} $\delta$~Lup -- two periods are claimed in the literature, however photometric period is likely an alias of the spectroscopic one (\citealt{ASGH05}); KK Vel -- two frequencies $f$ and $2f$ are detected in this variable. However, it is not clear if $2f$ is the harmonic frequency or an independent, nonlinearly coupled mode as suggested by \citet{CAea94}. Even in the latter case, KK Vel can be treated as a {\it monoperiodic} variable, similarly to the bump Cepheids.
\label{tabamp}
\end{table}

\section{Construction of model sequences}
\begin{figure*}
\includegraphics[width=176mm]{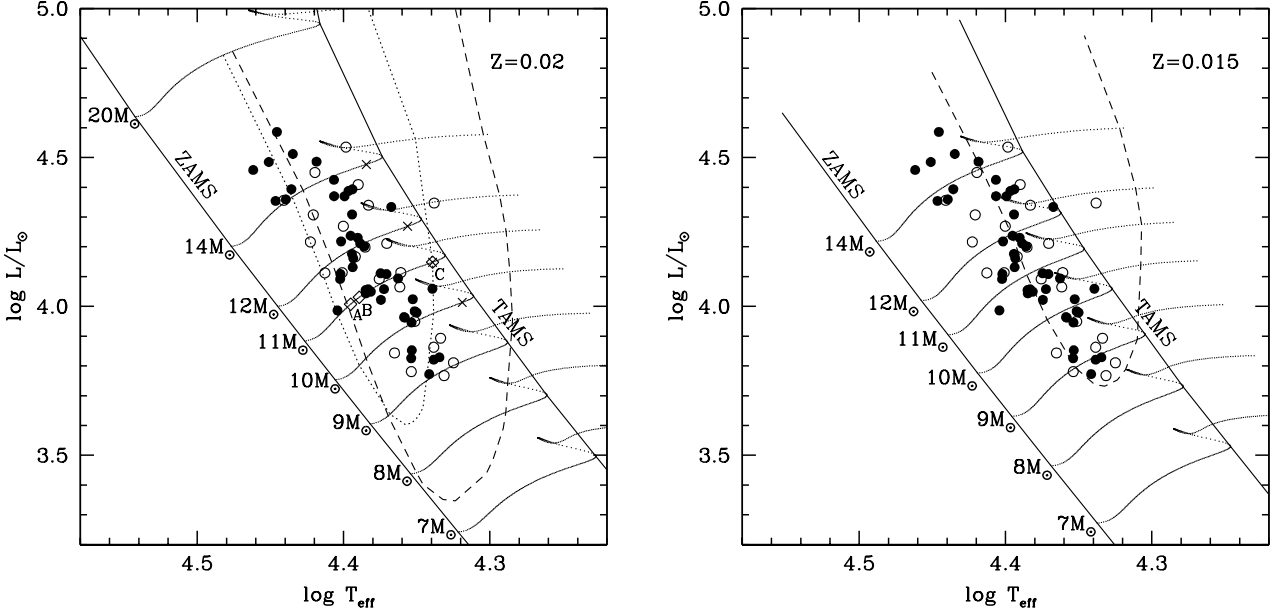}
\caption{Evolutionary tracks and LNA stability results for $\beta$~Cephei models with $Z=0.02$ (left panel) and $Z=0.015$ (right panel). Dashed and dotted lines enclose the fundamental and first overtone instability strips, respectively. Circles represent observational data: open circles -- monoperiodic variables, filled circles -- multiperiodic variables. Diamonds refer to positions of test models, A, B and C, described in Section~3, while crosses refer to convective models described in Section~4.4.}
\label{trak}
\end{figure*}

Evolutionary main-sequence models were computed using Warsaw-New Jersey evolutionary code (for details see e.g. \citealt{AP99}, and references therein). Rotation and overshooting from the convective core were disregarded. As will be justified in Section~4.4, convection in model envelopes may be neglected since $\beta$~Cephei stars are too hot to have an effective energy transfer by convection in stellar envelope. Several tracks with metallicities $Z=0.015$ and $Z=0.02$, and masses 7--14\thinspace M$_\odot$ were calculated. Additional high-mass track (20\thinspace M$_\odot$, $Z=0.02$) was also computed. The OPAL opacities were used in all evolutionary calculations. Computed evolutionary tracks are presented in Figure~\ref{trak}. For comparison we also plot galactic $\beta$~Cephei variables (circles in Fig.~\ref{trak}). Their bolometric luminosities and effective temperatures were obtained using Str\"omgren photometry indices collected in \citet{ASGH05} catalogue. Effective temperatures were obtained using calibrations of \citet{TMMD85} (\textsc{uvbybeta} code written by Moon and modified by Napiwotzki). $M_V$ and bolometric corrections were derived via calibrations of \citet{LBRS84} and from the work of \citet{PF96}, respectively. Monoperiodic variables are plotted as open circles. They do not form any distinct group.

Bolometric luminosities and effective temperatures, resulting from evolutionary calculations, were used in construction of model envelopes, that were subject to further radial pulsation analysis. Static envelopes were calculated using 120 zone mesh, with zone mass increasing geometrically inward. The temperature of the 70th zone from the surface was fixed to $2.5\times 10^{5}$K. The zone mass ratio above this anchor was the same for all models, while below the anchor it was iterated to match the desired bottom boundary temperature, $T_{\mathrm{in}}$, equal $3\times 10^{7}$K. Described mesh assures reasonable resolution in a shock region developing in exterior zones of non-linear models.

All models were subject to linear pulsation analysis and selected models were subject to non-linear calculations. We used radiative Lagrangean hydrocodes, suitable for radial pulsation anlysis. These codes are essentially those of \citet{RS75}, with modifications described in \citet{GKRB88}. In the linear non-adiabatic (LNA) code we used canned eigenvalue solver as suggested by \citet{AGRB93}. We assumed $\sim\exp (\sigma t)$ dependence of the perturbed quantities, where $\sigma$ is the complex eigenvalue. We define the linear growth rates of the modes as fractional growth of the kinetic energy per pulsation period: $\gamma=4\pi\Re (\sigma)/\omega$, where: $\omega=\Im (\sigma)$, is mode frequency. Analytical equation of state \citep{RS82}, and most recent version of OPAL opacities \citep{CIFR96} were used. Opacity tables were generated for solar mixture of \citet{NGAN93}. For some sequences OP opacities \citep{MS05} with Seaton mixture \citep{MSea94} were used. Linear stability results are presented in Figure~\ref{trak}. Fundamental and first overtone instability strips ($\gamma_0>0,\ \gamma_1>0$) are enclosed by dashed and dotted lines, respectively. First overtone is unstable for $Z=0.02$ only. Higher-order overtones are always stable. Observed instability strip runs approximately parallel to the Zero-Age Main Sequence (ZAMS) and the Terminal-Age Main Sequence (TAMS) lines. Theoretical instability strip extends to post main sequence phase, but since evolution becomes very fast shortly after the TAMS, the TAMS is often considered as the effective red edge of the pulsation instability domain.

For selected models, full amplitude non-linear limit cycles (periodic finite-amplitude pulsations) of fundamental and first overtone modes were calculated using relaxation scheme \citep{NBKS69,RS74}. The relaxation scheme provides also information about limit cycle stability (Floquet analysis, see e.g. \citealt{ELI56}). The same numerical mesh, difference scheme and opacity/equation of state subroutines were used both in LNA and in hydrocodes. Shocks were treated with an artificial viscosity formula of \citet{RS75}, with artificial viscosity parameters $C_Q=4$, and cutoff parameter $\alpha=0.06$. The choice of viscosity parameters will be discussed in Section~3.1. In non-linear iterations we used 400 time steps per pulsation period and in all cases it was enough to obtain reliable amplitude. However,  for evolved fundamental mode models with shocks, up to 2000 time steps per period were necessary to obtain reliable values of Floquet stability coefficients. Since most of the observational data are in a {\em V}-band, theoretical light curves were transformed to {\em V}-band through \citet{RK06} static atmosphere models. This was done by applying a bolometric correction at each pulsation phase, however omitting the shock phase, where this quasi-static approximation is not applicable.

\section{Sensitivity of calculated amplitudes}
Since conclusions of this paper rely on calculated limit cycle amplitudes, it is necessary to check how reliable they are. It is well known from classical pulsators modelling, that for a given model ($M$, $T_{\mathrm{eff}}$, $L$, $X$, $Z$), limit cycle saturation amplitudes may vary depending on artificial viscosity and numerical mesh \citep{GK90}. In this Section these effects are shortly analysed. All calculations are performed for bolometric light curves, since they are more smooth than {\em V}-band curves and void of transformation uncertainties (as we found, simple linear relation of type $A_V=a\cdot A_{\mathrm{bol}}$ exist, thus results concerning amplitude uncertainty are the same for {\em V}-band curves). Tests were done for the fundamental mode pulsations of three models, denoted by A, B and C in Fig.~\ref{trak}. All models have $M=11$\thinspace M$_\odot$, $X=0.7$ and $Z=0.02$. Models A and B are close to the boundary of the instability strip, while model C is an evolved model approaching the TAMS. In all calculations 400 time steps per pulsation period were used. Test calculations for other models and for the first overtone mode showed, that results presented below are representative.

\subsection{Sensitivity to artificial viscosity}
\begin{figure*}
\includegraphics[width=176mm]{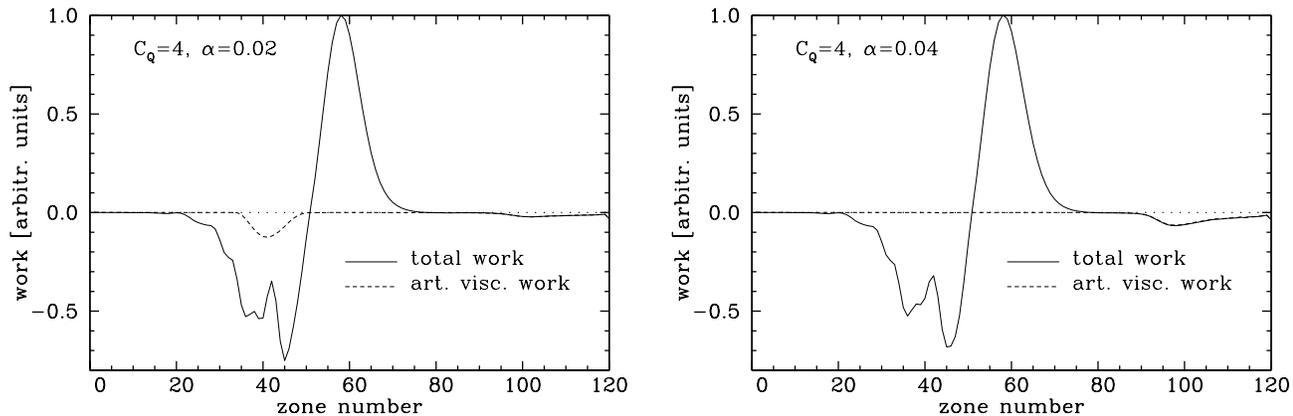}
\caption{Non-linear limit cycle work per zone for different artificial viscosity cutoff parameters: $\alpha=0.02$ (left panel) and $\alpha=0.04$ (right panel). Calculations performed for test model C (Fig.~\ref{trak}). Surface at right.}
\label{workintegrals}
\end{figure*}

\begin{figure}
\includegraphics[width=84mm]{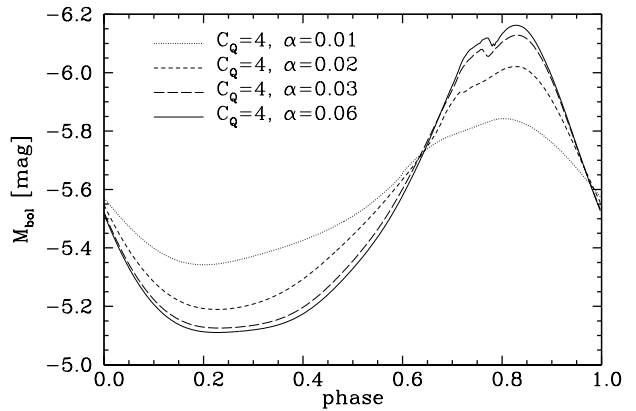}
\caption{Effects of different artificial viscosity parameters on bolometric light curve. Calculations performed for test model C (Fig.~\ref{trak}).}
\label{avcurves}
\end{figure}

In case of radiative models of classical pulsators, artificial viscosity parameters determine the limit cycle amplitude, and must be chosen to satisfy the observational constraints. Fortunately, this is not the case for $\beta$~Cephei stars. Calculations performed for test model C show, that for sufficiently high cutoff parameter ($\alpha>0.04$) unphysical interior dissipation, which affects the amplitude, is eliminated. Artificial viscosity turns on only in the exterior shock region, with almost no effect on limiting amplitude. This is illustrated through non-linear work integrals (Fig.~\ref{workintegrals}) and bolometric light curves (Fig.~\ref{avcurves}) obtained with different cutoff parameters. With $\alpha=0.02$ (left panel of Fig.~\ref{workintegrals}) one note significant negative contribution of artificial viscosity work in radiative damping zones (around zone 40), which leads to decrease of amplitude (Fig.~\ref{avcurves}). This dissipation is unphysical since no shock is generated here. Thus, we need to increase the cutoff parameter in order to eliminate this unphysical dissipation. For $\alpha=0.04$ (right panel of Fig.~\ref{workintegrals}) artificial viscosity contributes only in the exterior, shock zones. As it is well visible in Fig.~\ref{avcurves}, with increasing $\alpha$, amplitudes converge to constant value. Indeed, light curves obtained for $\alpha\in (0.04,\ 0.1)$ overlap and for clarity are not presented in Fig.~\ref{avcurves}. Also keeping $\alpha=0.06$ fixed and varying $C_Q$ parameter in a range, $C_Q\in\{2,\ 3,\ 4,\ 5,\ 6\}$, resulting light curves are indistinguishable. This justifies our choice: ($C_Q,\ \alpha)=(4,\ 0.06)$ and conclusion that for sufficiently high cutoff parameter, amplitudes are independent of artificial viscosity.

\subsection{Sensitivity to numerical mesh}
We found that amplitudes are almost insensitive to numerical mesh in case of evolved models, and only slightly sensitive for models in close vicinity of the blue edge of the instability strip. This is illustrated in Table~\ref{tabsen}. First two columns give mesh parameters: number of zones, $N$, and temperature of the innermost zone, $T_{\mathrm{in}}$. In the next columns there are bolometric light amplitudes and percentage amplitude difference in comparison to model with standard parameters ($N$=120, $T_{\mathrm{in}}=3\times 10^{7}$K). Stronger sensitivity of amplitudes to numerical mesh in direct vicinity of the blue edge is not surprising, since near the edge balance between damping and driving is a delicate function of our relatively coarse mesh. Test models with different location and value of anchor have amplitudes varying on similar level as presented in Table~\ref{tabsen}. The chosen depth of the envelope ($T_{\mathrm{in}}=3\times 10^{7}$K), seems sufficient to obtain reliable amplitudes. We also note that linear properties agree well with results obtained with more accurate linear code of Dziembowski \citep{WD77}. We conclude that for most of the main and post-main sequence models, limit cycle amplitudes are reliable to within 3 per cent.

\begin{table}
\caption{Bolometric peak-to-peak amplitude, $A_{\mathrm{bol}}$, sensitivity to mesh parameters. $N$ stands for number of zones and $T_{\mathrm{in}}$ correspond to inner temperature of the model envelope.}
\begin{tabular}{cccccccc}
\hline
    &          &\multicolumn{2}{c}{A}            & \multicolumn{2}{c}{B}  & \multicolumn{2}{c}{C}     \\
$N$ & $T_{\mathrm{in}}$ &$A_{\mathrm{bol}}$&        &$A_{\mathrm{bol}}$&         &$A_{\mathrm{bol}}$&        \\
\hline
120 & $3\times 10^{7}$  & 0.50    &  $-$   & 0.95    &  $-$    & 1.05    &  $-$   \\
160 & $3\times 10^{7}$  & 0.56    &  +12\% & 0.98    &  +3.2\% & 1.06    &  +1.0\%\\
120 & $4\times 10^{7}$  & 0.42    &$-$16\% & 0.93    &$-$2.1\% & 1.03    &$-$1.9\%\\
\hline
\end{tabular}
\label{tabsen}
\end{table}

\section{Results}
In this Section we present results of main survey (Section~4.1) and their interpretation (Section~4.2). In Section~4.3 we discuss the effects of different opacities and in Section~4.4 justify the neglection of convection.

\subsection{Main survey (OPAL opacities)}
\begin{figure*}
\includegraphics[width=176mm]{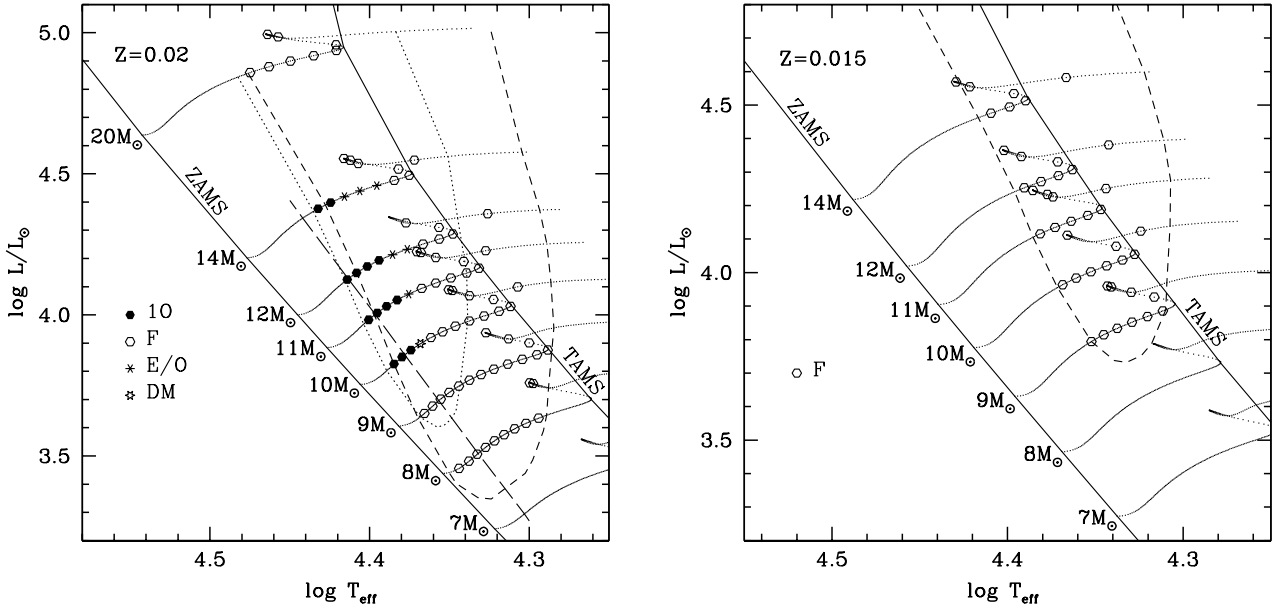}
\caption{Location of the calculated non-linear $\beta$~Cephei models, with modal selection information. 1O refer to the first overtone mode, F to the fundamental mode, E/O and DM correspond to either-or and double-mode behaviour, respectively. Left panel for $Z=0.02$ and right panel for $Z=0.015$. Long-dashed line in the left Figure shows the location of the $2\omega_1\simeq\omega_0+\omega_2$ resonance centre.}
\label{traksel}
\end{figure*}

\begin{figure*}
\includegraphics[width=176mm]{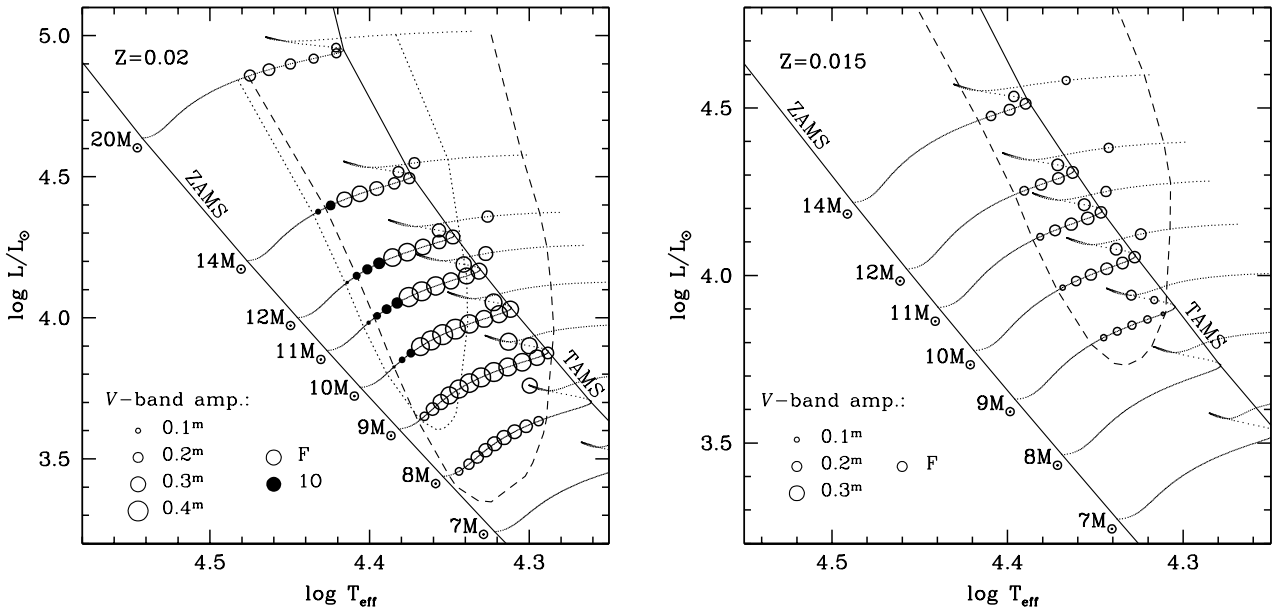}
\caption{{\em V}-band amplitudes of the non-linear models as a function of their location on the theoretical HR diagram. For E/O models and one DM model fundamental mode amplitude is plotted. Left panel for $Z=0.02$ and right panel for $Z=0.015$.}
\label{trakamp}
\end{figure*}

\begin{figure}
\includegraphics[width=84mm]{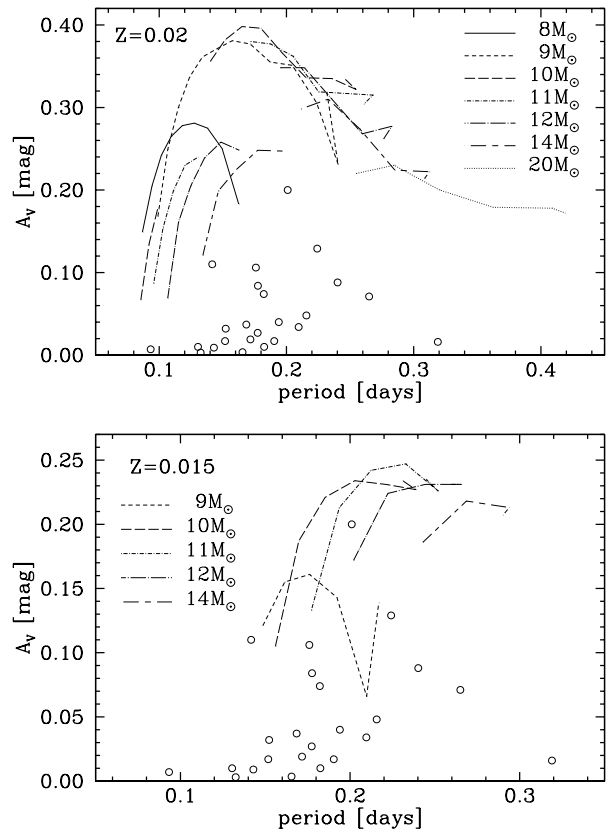}
\caption{{\em V}-band amplitudes of the non-linear models plotted against their period. In case of E/O models both the fundamental and first mode amplitudes are plotted. First overtone models (10--14\thinspace M$_\odot$) concentrate near periods of 0.14 days. Circles correspond to the observed, apparently monoperiodic $\beta$~Cephei stars. Upper panel for $Z=0.02$ and lower panel for $Z=0.015$.}
\label{ampper}
\end{figure}

Location of the calculated non-linear models on the theoretical HR diagram is shown in Figure~\ref{traksel} for $Z=0.02$ (left panel) and $Z=0.015$ (right panel). Since evolution becomes very fast shortly after the TAMS, the TAMS is considered as an effective red edge of the pulsation domain. Thus, only a few models are calculated beyond the TAMS. Figure~\ref{traksel} presents  modal selection information, obtained through linear analysis and Floquet analysis of the non-linear limit cycles. If only one mode is linearly unstable, then this mode is the only attractor of the system. If both modes are linearly unstable, selection scenario may be drawn from the Floquet analysis of the fundamental and first overtone limit cycles, which provides switching rates (non-linear growth rates, Floquet stability coefficients) toward the first overtone, $\gamma_{1,0}$, and toward the fundamental mode, $\gamma_{0,1}$, respectively. Generally, $\gamma_{i,j}$ are a measure of the linear stability of the limit cycle of mode $j$ to perturbation with respect to all the possible modes $i$. In all our calculations $\gamma_{i\ge 2,1}$ and $\gamma_{i\ge 2,0}$ are always negative, thus only the fundamental and first overtone modes are considered in the following. Assuming that model sequences are non-resonant (as will be discussed in Section~5, parametric resonance, $2\omega_1\simeq\omega_0+\omega_2$, play a role here, but only in a very narrow range of $\log T_{\mathrm{eff}}$), and may be described by cubic amplitude equations, four scenarios are possible (e.g. \citealt{WDGK84}). If $\gamma_{0,1}<0$ and $\gamma_{1,0}>0$ then the first overtone (1O) is the only attractor of the system. If reverse is true ($\gamma_{0,1}>0$ and $\gamma_{1,0}<0$) fundamental mode (F) is the only attractor. If both limit cycles are stable, $\gamma_{0,1}<0$ and $\gamma_{1,0}<0$, either fundamental or first overtone pulsation is possible (so called either-or, E/O, situation). Finally if both limit cycles are unstable, $\gamma_{0,1}>0$ and $\gamma_{1,0}>0$, star must undergo double-mode (DM) pulsations. It is clearly visible in Figure~\ref{traksel} that fundamental mode pulsation is dominant. First overtone pulsation is restricted to the vicinity of the blue edge. Between the fundamental and first overtone pulsation domains, E/O or DM (10\thinspace M$_\odot$) domains emerge. In case of $Z=0.015$ only fundamental mode pulsation is possible. In Figure~\ref{trakamp} we display the {\em V}-band peak-to-peak amplitudes as a function of model location on the HR diagram. For E/O models and one DM model, only the fundamental limit cycle amplitude is plotted. In Figure~\ref{ampper} amplitudes are plotted vs. the model period. Only main sequence models (and one model located directly after the TAMS) are included  in Figure~\ref{ampper}. For E/O models amplitudes of both limit cycles are plotted. Open circles in Figure~\ref{ampper} correspond to observed monoperiodic variables. Their periods and peak-to-peak {\em V}-light amplitudes are collected in Table~\ref{tabamp}. Several conclusions may be drawn from Figures~\ref{trakamp}~and~\ref{ampper}:
\begin{enumerate}
\item Amplitudes of higher metallicity models ($Z=0.02$) are larger than those of lower metallicity models ($Z=0.015$). It is a simple consequence of $\kappa$-mechanism acting in the metal opacity bump.

\item The highest amplitudes are obtained for intermediate mass models (10--11\thinspace M$_\odot$), while saturation amplitudes are smaller for less massive models (8--9\thinspace M$_\odot$) and there is a continuous fall of amplitudes for higher masses ($M>11$\thinspace M$_\odot$), supported by additional calculated sequence of 20\thinspace M$_\odot$ ($Z=0.02$). Near the TAMS and in the post-main sequence phase, saturation amplitudes are smaller than in the middle of main sequence band, but this fall of amplitudes is not very significant (Fig.~\ref{trakamp}).

\item The model amplitudes are higher than the observed ones (Fig.~\ref{ampper}). Considering rough averages of all the model and observed amplitudes (Table~\ref{tabamp}), the former ones are higher approximately by factor $\sim$4 and $\sim$6 in case of $Z=0.015$ and $Z=0.02$ model sequences, respectively. Calculating average amplitude for non-linear models we used main sequence models of masses 9--14\thinspace M$_\odot$, as these models correspond to observed variables (cf. Fig.~\ref{trak}). For E/O models and one DM model we chose the fundamental mode amplitudes (choosing first overtone amplitudes instead, has no effect on presented estimation).
\end{enumerate}
Calculated single mode saturation amplitudes are significantly higher than amplitudes of monoperiodic $\beta$~Cephei variables. As is clearly visible in Figures~\ref{trakamp}~and~\ref{ampper}, theoretical amplitudes may by lowered be decreasing metal abundance of the models. After lowering metal abundance from $Z=0.02$ to $Z=0.015$ theoretical amplitudes are still significantly higher than the observed ones. At the same time the instability strip shrinks and leaves a lot of stars beyond the blue edge, as is well visible in Figure~\ref{trak}. Lowering $Z$ further will increase this discrepancy. Thus, by lowering the metal abundance $Z$, we are not able to match simultaneously the observed amplitudes and the instability domain.

The natural explanation for described amplitude discrepancy is, that additional, undetected modes must be excited in {\em apparently} monoperiodic variables in order to account for the low amplitude of the only observed mode. Since these additional modes are not visible, either their amplitudes are very low or they are of high-degree $l$, and are not visible due to averaging. In the following we show that the collective instability saturation by acoustic modes is sufficient to explain the amplitudes observed in $\beta$~Cephei variables.

\subsection{Collective saturation of the pulsation instability}

Collective saturation of the pulsation instability may be described by amplitude equations. For each $i$-th mode of the $n$ non-resonantly interacting modes amplitude changes are given by (e.g.~\citealt{WDGK84}):
\begin{equation}
\frac{dA_i}{dt}=\gamma_i\big(1-\alpha_{i1}A_1^2+\ldots-\alpha_{in}A_n^2\big)A_i
\label{ae}
\end{equation}
where $\alpha_{ij}$ are saturation coefficients. Pulsation instability is saturated when all $dA_i/dt=0$. Thus, it is clear, that collective saturation of the pulsation instability by several acoustic modes will lead to amplitudes lower than in single mode case. Quantitative estimates may be drawn assuming that all acoustic modes are similar, that is have the same saturation coefficients, $\alpha_{ij}\equiv \alpha$. Then, a simple consequence of equations~(\ref{ae}) is that pulsation instability may be saturated either by a single mode with amplitude $A=1/\sqrt{\alpha}$ or by $n$ simultaneously excited modes, each of amplitude $A_n=1/\sqrt{n\alpha}$ (thus, lower by a factor $\sqrt{n}$ than in single mode case). In the following, we consider average model and average monoperiodic $\beta$~Cephei variable. As was already discussed in the previous section, for $Z=0.02$ model amplitudes are $\sim$6 times higher than observed amplitudes ($\sim$4 times for $Z=0.015$). Thus with $\sim$36 modes of the same amplitude as is observed in average monoperiodic variable, one is able to saturate the pulsation instability (with $\sim$16 modes in case of $Z=0.015$ models). Of course these are lower limits for number of modes, since in this hypothetic multiperiodic variable all the modes should be visible. We stress that these are rough estimates, since saturation coefficients of different modes do differ. However, it is clear that if collective saturation is responsible for amplitude limitation a dozen or so modes must be available (linearly unstable). In order to match the amplitudes, collective saturation  also requires higher number of excited modes for more metal rich variables. Fortunately this is the case as linear theory shows \citep{WDAP93}. We also note that for observed multiperiodic $\beta$~Cephei variables, higher number of detected modes is correlated with higher metallicity \citep{JDEN05}.

\begin{figure}
\includegraphics[width=84mm]{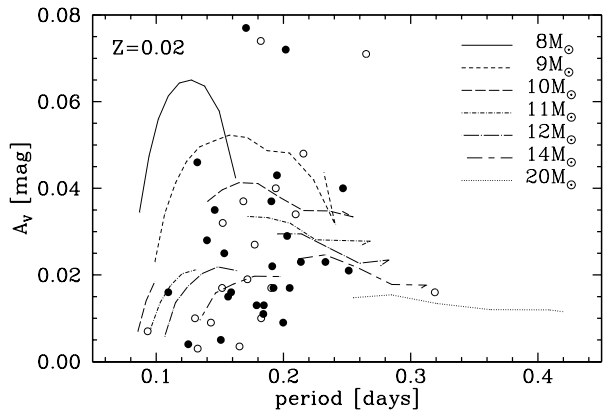}
\caption{Theoretical amplitudes, recalculated under assumption of collective saturation of the instability mechanism by 1/3 of linearly unstable acoustic modes. In addition to monoperiodic $\beta$~Cephei variables (open circles) also multiperiodic are plotted (filled circles).}
\label{ampperres}
\end{figure}

To check whether a dozen or so non-radial modes are available (linearly unstable) we used LNA code of \citet{WD77}. In this code, full evolutionary stellar model is subject to linear pulsation analysis. We determined the number of linearly unstable acoustic modes for models of different masses ($Z=0.02$), located in the centre of the main sequence band. We considered modes of degree $l<17$ and dimensionless frequency, $\sigma$ ($\sigma=\omega/\sqrt{4\pi G\bar{\rho}}$) in a range $1<\sigma<5$. This choice is arbitrary, but well separates groups of acoustic, gravity and mixed modes (separation is clear for lower masses, and worsens for higher). Results are presented in~Table~\ref{tabmod}, where number of unstable modes, taking into account rotational splitting, is given. One should look at orders of presented values and trends among them, rather than at specific values, since single additional, high degree mode, may affect the results presented in the Table. As we have checked, the number of unstable modes doesn't vary much along evolutionary track. However, the trend of increasing number of modes with increasing mass is clearly visible. It is clear that considering only acoustic modes, more than enough modes are available for collective instability saturation. We note that not all linearly unstable modes must take part in the saturation, as was shown in case of $\delta$~Scuti models \citep{RN05}. The number of linearly unstable modes is also much higher, than observed in multiperiodic $\beta$~Cephei stars. Thus, assuming that each number given in Table~\ref{tabmod} is representative for entire evolutionary track of a given mass and that only one third of these modes take part in the saturation process, we rescaled theoretical single mode amplitudes (Figure~\ref{ampper}), under assumption of collective saturation. Results are presented in Figure~\ref{ampperres}. Using only part of the linearly unstable acoustic modes, we lowered predicted amplitudes to the observed level. Thus, we argue that collective instability saturation is sufficient to explain amplitude limitation in the $\beta$~Cephei pulsators.

\begin{table}
\begin{tabular}{cc}
$M$ & $n$ \\
\hline
 8\thinspace M$_\odot$  &  56\\
 9\thinspace M$_\odot$  & 159\\
10\thinspace M$_\odot$  & 278\\
11\thinspace M$_\odot$  & 385\\
12\thinspace M$_\odot$  & 418\\
14\thinspace M$_\odot$  & 475\\
20\thinspace M$_\odot$  & 665\\
\hline
\end{tabular}
\caption{Results of detailed linear analysis of selected models. $n$ stands for the number of unstable modes considering rotational splitting.}
\label{tabmod}
\end{table}

A closer look at Figure~\ref{ampperres} shows that the strongest reduction of amplitudes occurs for 20\thinspace M$_\odot$ track. However, predicted amplitudes are still above current detection limit. Thus, under our simplifying approximations, non-linear effects cannot explain lack of massive and luminous $\beta$~Cephei pulsators. Fortunately, three candidates for such $\beta$~Cephei pulsators were found very recently \citep{AP07}. We note in passing that evolved massive pulsators of $\alpha$~Cygni type are known to occur in the upper part of the $\beta$~Cephei domain \citep{KLea07}. They are located near the TAMS and beyond the TAMS. Their variability is interpreted by \citet{KLea07} in terms of g-mode pulsation. As such, these varaibles are not genuine $\beta$~Cephei pulsators, but rather high-mass analogues of the SPB stars. Our calculations, which are directly applicable only to p-mode pulsators cannot be used to interpret amplitudes of these stars.

The lack of pulsators at the low luminosity end of the theoretical instability strip, can not be explained by non-linear effects, either. Even under assumption of collective saturation, predicted amplitudes are well above the detection limit. However, we stress that these conclusions rely on a simplifying assumption about similar properties of all acoustic modes. This simplification is useful, since it allows for some quantitative conclusions, but in real star saturation coefficients of modes do differ. One should remember that transformation from observational to theoretical HR diagram is uncertain. Possible systematic error in temperature transformations may explain lack of less massive pulsators near the TAMS, and presence of a few pulsators of higher mass beyond the blue edge (see Figure~\ref{trak}). Also using a new solar mixture of \citet{MAea05} partly solves these problems as it leads to theoretical instability strip more consistent with the observational one \citep{APWZ07}.

As we have shown, collective instability saturation is quite
sufficient to explain the observed amplitudes of $\beta$~Cephei
variables. A possible drawback of this mechanism is that
excitation of many high-$l$ modes results in broadening of
spectral lines, which might be too high when compared to
observations. This problem has already been discussed by
\citet{WD80} in context of $\delta$~Scuti variables. Here we
follow his reasoning.

Excitation of a pulsation mode induces time-dependent broadening
and distortion of spectral lines. The amplitude of this effect is
given, roughly, by amplitude of the photospheric velocity
variations. In case of collective instability saturation,
time-dependent contributions from hundreds of modes with different
periods add up to form a line profile. This would mimic
macroturbulence and would contribute to the measured width of the
spectral line. Exact modeling of this "collective" line-broadening
is a very laborious task and is beyond the scope of this paper.
Here, we estimate this broadening by estimating the root mean
square {\it macroturbulence velocity} induced by all pulsation
modes involved. Again, we assume that all modes have the same
saturation coefficients and that they can be treated as
statistically independent. Single radial mode creates photospheric
velocity field $v_0(t)$ with amplitude $V$, which is given by our nonlinear calculations. $V$ of order of 100 km s$^{-1}$ is typical for our models. When pulsation instability is
saturated by $n$ acoustic modes, amplitudes of individual modes
are a factor of $\sqrt{n}$ lower, namely $v_n(t) \simeq
V/\sqrt{n}\cos\omega_n t$ (see previous discussion). The root mean
square macroturbulence velocity can be estimated by simple
averaging:
\begin{equation}
\sqrt{\big<\sum v_n^2(t)\big>} =
\sqrt{\sum\frac{V^2}{n}<\cos^2\omega_n t>} = \frac{V}{\sqrt{2}}
\end{equation}
\noindent Thus, on the basis of our nonlinear calculations, we
expect the line broadening in collective saturation scenario to be
of the order of 100 km s$^{-1}$. This number is, generally, comparable to
$v\sin i$ values measured in $\beta$~Cephei variables
(\citealt{ASGH05}, fig.~2), however, for many stars the observed
line broadening is smaller. This poses a potential problem to the
proposed scenario. A possible way out of this difficulty is to
include the g-modes (omitted in our analysis) into the saturation
process. $\beta$~Cephei model calculations kindly provided by
Dziembowski (private comm.) show, that photospheric velocity
variations of the g-modes should be smaller than those of the
p-modes. This is true, assuming the same flux perturbation
amplitudes in the driving zone, which is the natural normalisation
when instability saturation is discussed.

\subsection{Effects of different opacities}
\begin{figure}
\includegraphics[width=84mm]{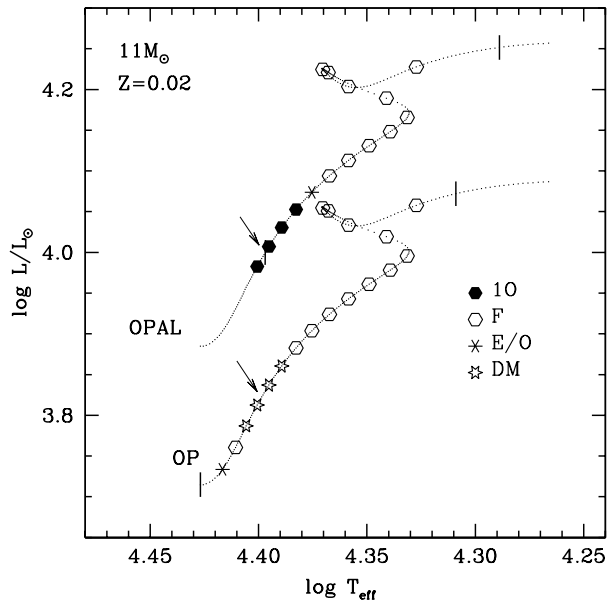}
\caption{11\thinspace M$_\odot$ model sequence calculated using both the OPAL and OP opacities in the pulsation codes. Since evolutionary track was calculated using the OPAL opacities, track with the OP results is artificially shifted to allow comparison. Vertical ticks mark the linear edges of the fundamental mode instability strip. Arrows indicate the position of the $2\omega_1\simeq\omega_0+\omega_2$ resonance centre.}
\label{opacsel}
\end{figure}

\begin{figure}
\includegraphics[width=84mm]{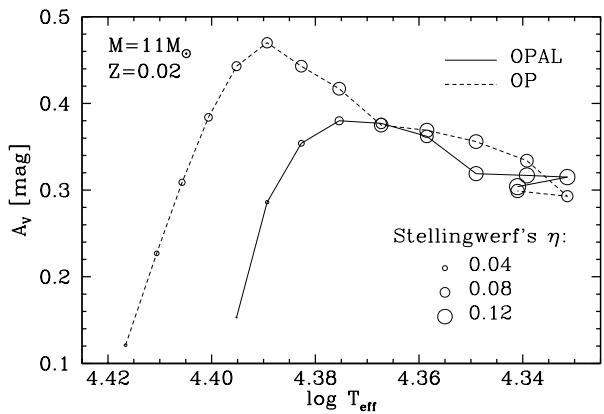}
\caption{The {\em V}-band peak-to-peak fundamental mode amplitudes of 11\thinspace M$_\odot$ models calculated with both the OPAL (solid line) and OP (dashed line) opacities. Size of the circles correspond to model's Stellingwerf's $\eta$.}
\label{opacamp}
\end{figure}

Linear properties of the $\beta$~Cephei models are sensitive to opacities being used. Particularly, the OP opacities lead to wider instability strip, as demonstrated by \citet{AP99} (during preparation of this paper, we have noticed that also different methods of interpolation of the OPAL opacities, lead to slightly different results, but since this effect is small, and has no effect on presented conclusions, we won't discuss it). We calculated one sequence of models ($M=11\thinspace \mathrm{M}_\odot$, $Z=0.02$) using the OP opacities. Models have the same $\log L$ and $\log T_{\mathrm{eff}}$ parameters as models using the OPAL opacities, since we are interested in opacity effect on pulsation properties. Models with modal selection  information are presented in Figure~\ref{opacsel}. To allow comparison, evolutionary track with the OP results is artificially shifted. Vertical ticks mark the edges of the fundamental mode instability strip. One may observe completely different modal selection scenario, with wide region of DM models for the OP sequence. Since robust DM behaviour, in purely radiative models, was not encountered up to date (cf. however \citealt{GKRB93}), we will discuss it in Section~5, here concentrating on amplitudes only. Figure~\ref{opacamp} presents peak-to-peak {\em V}-band amplitudes of the fundamental limit cycle as a function of model's effective temperature. Size of the circles correspond to \citet{RS78} linear growth rate, $\eta$. It characterises the envelope driving: with $\eta=1$ the entire envelope drives, while with $\eta=-1$ the entire envelope is dissipative. Amplitudes obtained with the OP opacities, are generally higher, which is connected with higher $\eta$ values. However for temperatures higher than $\log T_{\mathrm{eff}}\approx 4.37$ visible difference between the OP and OPAL results is connected with the location of the blue edge, significantly shifted to the blue in case of the OP opacities. For lower temperatures amplitudes obtained with the OP opacities are comparable to amplitudes obtained using the OPAL opacities. Thus, presented conclusions remain unchanged.

\subsection{Effects of convection}
Contrary to the classical pulsators, convection is not expected to play significant role in hot $\beta$~Cephei stars. As we checked for 11\thinspace M$_\odot$, evolutionary track calculated with mixing-length convection ($\alpha=1$) overlap the track calculated without convection (see also \citealt{AP99}). Impact of convection on pulsation models was checked using direct time integration convective code. We used Lagrangean code with time-dependent convection model of \citet{RK86} in the version of \citet{GWMF98}. All other hydrodynamics and procedures are the same as were used in radiative hydrocodes. Since growth rates of $\beta$~Cephei stars are small, convergence to limit cycle may require calculation of more than $10^5$ pulsation cycles. For this reason only four models and two sets of convective parameters were checked ($\alpha\in\{1.0,\ 0.5\}$, $\alpha_m=0.4$; $\alpha_s$, $c_D$, $\gamma_r$ -- standard values as given by \citealt{GWMF98}). Turbulent pressure and overshooting were neglected ($\alpha_p=\alpha_t=0$). Tested models are evolved and massive models pulsating in the fundamental mode. These models are characterised by relatively high growth rates (limiting amplitude is achieved faster) and have stable first overtone (transient phase is short). Models are marked by crosses in Figure~\ref{trak}. As initial model for integration we used fully consistent static convective envelope, constructed in the same way (number of zones, anchor, depth of the envelope) as for radiative models. We find that during the contraction phase up to 10 per cent of the flux may be carried by metal-bump induced convective zone. Limiting pulsation amplitudes are lower than in case of radiative models, but the difference is generally small, which is shown in Table~\ref{tabcon}. Lower amplitudes for convective models are expected, since part of the flux is carried by convection and thus the $\kappa$-mechanism is less effective. We note up to 18 per cent difference between radiative and $\alpha=1$ convective amplitudes for less massive models. With $\alpha=0.5$ convective and radiative amplitudes are almost the same. We argue that $\alpha=0.5$ is a better value, since the mixing-length should be smaller than the extent of the convective zone. This condition is satisfied better with smaller $\alpha$ values.  However, we stress that even with $\alpha=0.5$, at some pulsation phases the extent of the convective zone is comparable with the mixing-length. Thus, this model of convection, working well in case of classical pulsators, is at the limits of its applicability here and is used only for estimation of the impact of neglecting convective energy transfer on calculated amplitudes. We also expect that convective effects get weaker as one moves from the TAMS (where our test models are located) toward the ZAMS. Construction of the static convective envelopes for 10\thinspace M$_\odot$ evolutionary track clearly shows this. For this sequence maximal static convective flux for model presented in Table~\ref{tabcon} is 2.7 per cent of the total flux, while for the model closest to the ZAMS it is less than 0.4 per cent. Of course dynamical effects lead to higher values of convective flux in non-linear models, but we expect the same trend. We conclude, that although amplitudes of convective models are systematically lower, this difference is not significant for our discussion.

\begin{table}
\begin{tabular}{cccc}
$M$ & $A_{\mathrm{bol}}^r$ & \multicolumn{2}{c}{$A_{\mathrm{bol}}^c$}\\
    &                      & $\alpha=1$           & $\alpha=0.5$\\
\hline
14\thinspace M$_\odot$ & 0.85  & 0.83 & 0.85\\
12\thinspace M$_\odot$ & 0.98  & 0.95 & 0.98\\
11\thinspace M$_\odot$ & 1.05  & 0.99 & 1.05\\
10\thinspace M$_\odot$ & 1.16  & 0.98 & 1.14\\
\hline
\end{tabular}
\caption{Comparison of bolometric light peak-to-peak amplitudes of selected radiative ($A_{\mathrm{bol}}^r$) and convective ($A_{\mathrm{bol}}^c$) models computed with mixing-length equal 1 (third column) and 0.5 (fourth column).}
\label{tabcon}
\end{table}

\section{Double-Mode models}
Although DM RR~Lyrae and Cepheids are quite common, for many years purely radiative codes failed to reproduce the DM behaviour. In case of RR~Lyrae models \citet{GKRB93} showed, that for particular range of artificial viscosity parameters, the DM state is possible, but still their models were strongly dependent on numerical details, and were not in satisfactory agreement with observations. It was the inclusion of convection, that finally led to robust and satisfactory DM models \citep{MF98,ZKea98}. To our surprise we encountered one DM model in our radiative models built with the OPAL opacities (Fig.~\ref{traksel}), and a wide region of DM models in radiative models that use the OP opacities (Fig.~\ref{opacsel}). This was unexpected, since among observed multimode $\beta$~Cephei stars, pulsators with two radial modes only have not been detected so far \citep{CAPDC03}. Linear analysis showed, that despite obvious role of different opacities, also a parametric resonance, $2\omega_1\simeq\omega_0+\omega_2$, may be connected with DM phenomena. The resonance centre is indicated by the long-dashed line in Figure~\ref{traksel} and by arrows in Figure~\ref{opacsel}. As will be shown in the following, different mechanisms are responsible for DM-behaviour in case of the OPAL and the OP opacities and thus, models computed using both opacities are discussed in separate subsections. We also stress that our modal selection analysis relies only on Floquet stability coefficients (switching rates). If stability coefficients of both the fundamental and the first overtone limit cycles are positive, neither limit cycle is stable, and DM behaviour is unavoidable. However, only direct time integration provides information about amplitude stability of such solution. Due to very small growth rates, integration covering several hundred thousand cycles would be required. Such a calculation is prohibitively time-consuming and was not performed here.
\subsection{The case of the OPAL opacities}
It is clearly visible in Figure~\ref{traksel}, that the only DM model found in the main survey (10\thinspace M$_\odot$), is quite far from the resonance centre and thus, is not connected with it. It also seems, that the resonance does not lead to any change in the modal selection. However, this is not the case. We have calculated additional models of 10\thinspace M$_\odot$ in direct proximity of the resonance centre, and close to the only DM model visible in Figure~\ref{traksel}. Stability results are presented in Figure~\ref{rezopal}. Growth rates (both linear and non-linear) are plotted against parameter $\Delta$, defined as $\Delta=2\omega_1/(\omega_0+\omega_2)$. $\Delta$ is calculated using linear frequencies. The shaded bar indicates modal selection. Two DM domains emerge, with two different mechanisms being responsible:
\begin{enumerate}
\item {\bf Resonant DM.} The most striking feature visible in Figure~\ref{rezopal} is the prominent peak of switching rate toward the fundamental mode, $\gamma_{0,1}$, with no doubt connected with the resonance. Otherwise negative, $\gamma_{0,1}$ becomes positive at the resonance centre. Thus, the discussed resonance leads to the destabilization of the overtone limit cycle. As a result, the DM {\em island} appears in the middle of the first overtone pulsation domain. The domain of the resonant DM models is very narrow ($\Delta T_{\mathrm{eff}}\approx 200$K), and thus, no such models are visible in Figure~\ref{traksel}, since they fall in between the consecutive first overtone models. We also checked that for 11\thinspace M$_\odot$ evolutionary track there are DM models in direct proximity of the resonance, very close to the linear fundamental mode blue edge. For the 12\thinspace M$_\odot$ evolutionary track, resonance centre falls in between the linear edges of the fundamental and first overtone modes and the overtone limit cycle is stable ($\gamma_{0,1}<0$). Thus, the resonant DM models are restricted to intermediate masses, since for masses slightly above 12\thinspace M$_\odot$ resonance centre falls beyond the instability strip, while for smaller masses, below $\sim$9\thinspace M$_\odot$, $\gamma_{1,0}<0$, and thus the fundamental mode is the only stable attractor (Fig.~\ref{traksel}).
\item {\bf Non-resonant DM.} The second DM domain, at $\Delta\approx 1.01$ is not connected with any resonance. DM domain separates the first overtone and fundamental mode pulsation domains. Such behaviour is expected in case of non-resonant coupling of pulsation modes (e.g. \citealt{WDGK84}). The non-resonant DM behaviour is also restricted to intermediate masses ($\sim$10\thinspace M$_\odot$), since it requires that the switching rates do intersect, and are positive at the intersection. For less massive models the switching rates do not intersect, while, for massive models, intersection occurs at the negative value of switching rates, which leads to E/O behaviour.
\end{enumerate}

\subsection{The case of the OP opacities}

\begin{figure}
\includegraphics[width=84mm]{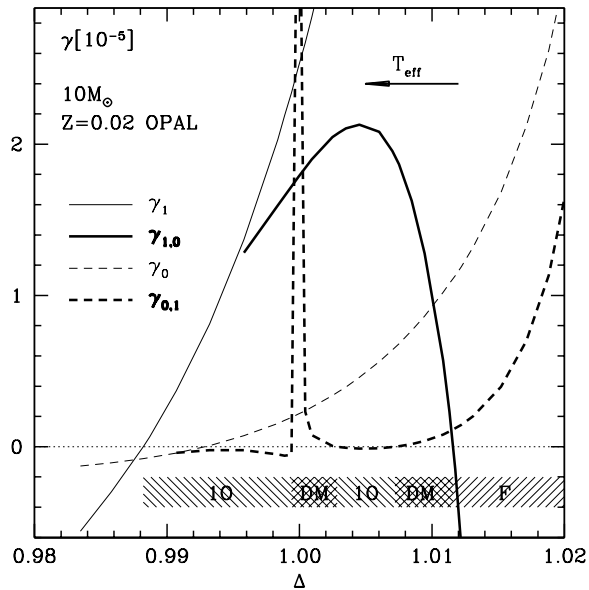}
\caption{Linear growth rates (thin lines) and non-linear growth rates (switching rates, thick lines) plotted against $\Delta$, $\Delta=2\omega_1/(\omega_0+\omega_2)$, for 10\thinspace M$_\odot$ evolutionary track computed using the OPAL opacities in pulsation calculations. The resonance centre ($\Delta=1$) is located at $\log T_{\mathrm{eff}}\approx 4.378$.}
\label{rezopal}
\end{figure}

\begin{figure}
\includegraphics[width=84mm]{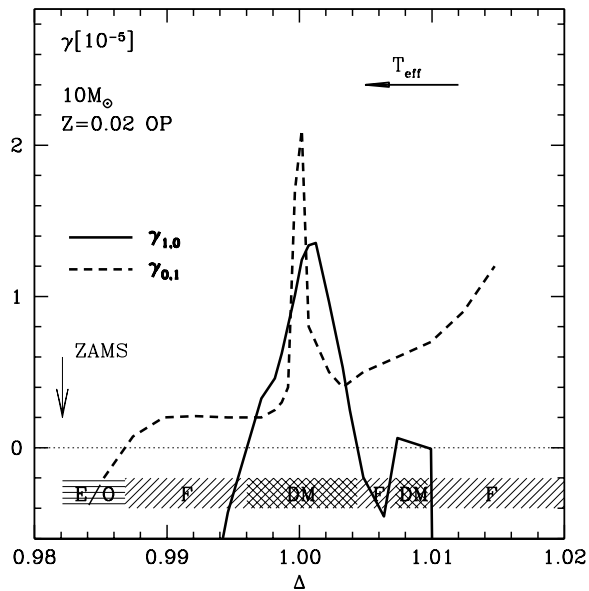}
\caption{Non-linear growth rates (switching rates, thick lines) plotted against $\Delta$, $\Delta=2\omega_1/(\omega_0+\omega_2)$, for 10\thinspace M$_\odot$ evolutionary track computed using the OP opacities in pulsation calculations. Arrow marks the location of the ZAMS. Both linear growth rates are positive almost in the entire range of $\Delta$ (both growth rates are positive at the ZAMS and the red edge of the first overtone pulsation falls at $\Delta\approx 1.018$) and therefore are not plotted. The resonance centre ($\Delta=1$) is located at $\log T_{\mathrm{eff}}\approx 4.379$.}
\label{rezop}
\end{figure}

\begin{figure}
\includegraphics[width=84mm]{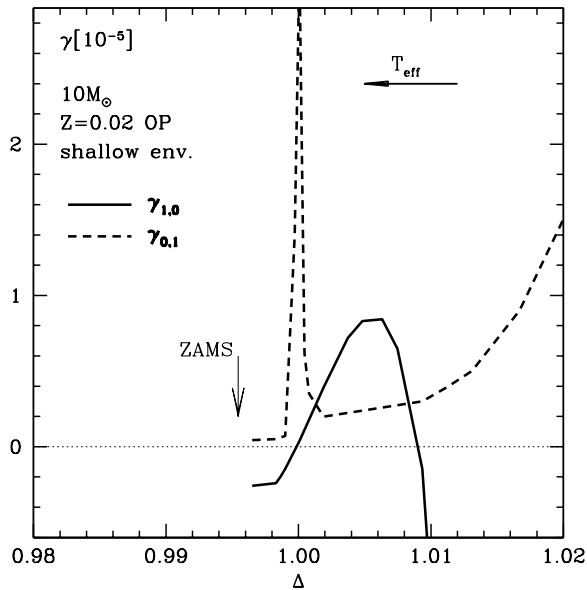}
\caption{Non-linear growth rates (switching rates) plotted against $\Delta$, $\Delta=2\omega_1/(\omega_0+\omega_2)$, for 10\thinspace M$_\odot$ evolutionary track computed using the OP opacities in pulsation calculations and shallow envelope model. Arrow marks the location of the ZAMS. Resonance centre is located at $\log T_{\mathrm{eff}}=4.392$.}
\label{rezenv}
\end{figure}

In case of the OP opacities, the domain of the DM-behaviour is much wider than in case of the OPAL opacities, as is well visible in Figure~\ref{opacsel}. It is also visible that the DM models seem to be concentrated near the resonance centre. However we will show that the resonance does not play any role here. To allow direct comparison with the OPAL results (Figure~\ref{rezopal}) we calculated several models of 10\thinspace M$_\odot$ in the proximity of the resonance. Switching rates are presented in Figure~\ref{rezop} (qualitatively the same picture is true for 11\thinspace M$_\odot$ evolutionary track for which modal selection results are presented in Figure~\ref{opacsel}). Both linear growth rates are positive (the red edge of the first overtone pulsation falls at $\Delta\approx 1.018$), and for clarity are not shown in the Figure. The prominent peak of $\gamma_{0,1}$ is well visible, but the resonance producing this peak is not responsible for the destabilization of the overtone limit cycle since, opposite to the OPAL case, $\gamma_{0,1}>0$ almost everywhere. Thus, it is the switching rate toward the first overtone, $\gamma_{1,0}$ (i.e. stability of the fundamental limit cycle) that determines the edges of the DM domain. $\gamma_{1,0}$ seems to peak near the resonance centre and there is a secondary maximum at around $\Delta\approx 1.008$. This second DM domain is probably the numerical artefact (probably connected with the strong shock propagation). We argue that the first DM domain falls at the $\Delta\approx1$ by accident, and thus $2\omega_1\simeq\omega_0+\omega_2$ resonance is not responsible for the destabilization of the fundamental limit cycle. Our arguments are as follows:
\begin{enumerate}
\item \citet{GKRB93} analysed the $2\omega_1\simeq\omega_0+\omega_2$ resonance, using lowest non-trivial order of the amplitude equations. They showed that the resonance affects only the stability of the overtone limit cycle, $\gamma_{0,1}$, while it doesn't affects the stability of the fundamental limit cycle, $\gamma_{1,0}$.
\item The same may be shown by numerical experiment. To check the effect of the resonance on stability coefficients, we calculated additional sequence of 10\thinspace M$_\odot$, with shallower envelope, inner temperature being, $T_{\mathrm{in}}=2.4\times 10^7$K. Resulting mode frequencies are slightly different in comparison to models with standard envelope and thus, the resonance centre is shifted to higher effective temperatures. Stability results are presented in Figure~\ref{rezenv}. The prominent peak of $\gamma_{0,1}$ is still well visible at the resonance centre. However the maximum of $\gamma_{1,0}$ has nothing to do with the resonance.
\end{enumerate}
We note that even if one treats the second DM domain visible in Figure~\ref{rezop} as a numerical artefact, the remaining picture is far from the one expected in case of non-resonant mode coupling, where DM domain separates the first overtone and fundamental mode pulsation domains \citep{WDGK84}. Here, the double-mode domain appears as an island in between the fundamental mode pulsation domains. We are not able to give a specific reason for the destabilization of the fundamental mode visible in Figure~\ref{rezop}.

\section{Conclusions}

In the present analysis we have focused on amplitude saturation in
$\beta$~Cephei models. We calculated several evolutionary tracks
of two different metallicities, $Z=0.02$ and $Z=0.015$, and masses
covering the observational instability domain. More than hundred
models were subject to non-linear pulsation analysis. Radial
modes, treated as representative for all acoustic oscillations,
were converged to limit cycle using \citet{RS74} relaxation
technique. We have shown that our models are robust, independent
of numerical treatment. Particularly, contrary to the classical
pulsators models, for physically correct choice of artificial
viscosity parameters the saturation amplitudes and light-curve
shapes of our models are independent of artificial viscosity.

Examination of saturation amplitudes along the model sequences, as
well as comparison of these amplitudes with amplitudes observed in
$\beta$~Cephei stars leads to several conclusions. Most important
concern amplitude limitation mechanism in $\beta$~Cephei stars.
Single mode saturation amplitudes are significantly higher than
amplitudes observed for monoperiodic $\beta$~Cephei variables.
Assuming similarity of all the acoustic modes, we showed that
collective instability saturation easily lowers predicted
amplitudes to the observed level. The straightforward conclusion
is, that in fact all $\beta$~Cephei pulsators should be
multiperiodic. Ongoing decrease of detection limit should reveal
additional modes in monoperiodic $\beta$~Cephei variables. This is
fully supported by recent progress in intensive multisite
campaigns leading to the detection of more and more pulsation
modes \citep{GHea04,GHea05,GHea06}. A possible weakness of the
collective saturation scenario is that the estimated
line-broadening induced by excitation of hundreds of pulsation
modes might be higher than observed in some stars. We argue that
this difficulty can be solved by allowing g-modes to take part in
the saturation. We also addressed the problem of non-uniform
filling of the theoretical instability strip, specially the lack
of massive and luminous p-mode pulsators. Although our
calculations show decrease of saturation amplitudes while moving
to higher masses, we found this effect too weak to explain
scarcity of such pulsators.

We have found that our results depend somewhat on the opacities
being used (OPAL or OP). Differences are significant but do not
change our conclusions. A few non-linear models were calculated
with turbulent convection based on \citet{RK86} theory. In
agreement with expectations, we found that convection may be
neglected in $\beta$~Cephei models.

In several of our radiative models we have found a double-mode
behaviour, with radial fundamental and first overtone modes
simultaneously excited. This form of pulsation is encountered only
in intermediate mass models (10--11\thinspace M$_\odot$). It is
numerically very robust and occurs both with OPAL and with OP
opacities. Depending on the specific model, the origin of double
mode pulsation can be traced to one of two different mechanisms:
either to the nonresonant coupling of the two excited modes (e.g.
\citet{WDGK84}), or to the $2\omega_1 \simeq \omega_0 + \omega_2$
parametric resonance.

Admitedly, our radial double-mode models have no direct
application to real stars, because no purely radial multimode
$\beta$~Cephei pulsators are known. Nevertheless, our models
demonstrate how easily low-order parametric resonances can
destabilize a single-mode radial pulsation. When all linearly
unstable (or marginally stable) {\it non-radial} modes are taken
into account, many more resonances of this type can be found. This
might explain why high ampliude monoperiodic radial pulsators
(similar to BW~Vul) are so rare among $\beta$~Cephei stars.

\section*{Acknowledgments}

We are grateful to prof. W. Dziembowski for many critical remarks and comments on the manuscript. We also thank W. Dziembowski and A. Pamyatnykh for the assistance while using Dziembowski's pulsation code and stellar evolution code. R. Napiwotzki and M. Jerzykiewicz are acknowledged for the \textsc{uvbybeta} code. This research has made use of the SIMBAD database operated at CDS Strasbourg, France. This work has been supported by the Polish MNiI Grant No. 1 P03D 011 30.

\bsp

\label{lastpage}

\end{document}